\documentclass[sigconf, nonacm]{acmart}

\usepackage{multicol}
\usepackage{csquotes}
\usepackage[utf8]{inputenc}
\usepackage{textgreek}

\AtBeginDocument{%
  }

\setcopyright{acmlicensed}
\copyrightyear{2025}
\acmYear{2025}
\acmDOI{XXXXXXX.XXXXXXX}
\acmConference[CHI '26]{The ACM Conference on Human Factors in Computing Systems}{April 13--17,
  2026}{Barcelona, Spain}

\acmISBN{978-1-4503-XXXX-X/2018/06}

\begin{document}

\title{Future You: Designing and Evaluating Multimodal AI-generated Digital Twins for Strengthening Future Self-Continuity}

\settopmatter{authorsperrow=4}

\author{Constanze Albrecht}
\authornote{These authors contributed equally to this work.}
\affiliation{%
  \institution{MIT Media Lab}
  \city{Cambridge}
  \state{MA}
  \country{USA}
}
\email{csophie@mit.edu}

\author{Chayapatr Archiwaranguprok}
\authornotemark[1]
\affiliation{%
  \institution{MIT Media Lab}
  \city{Cambridge}
  \state{MA}
  \country{USA}
}
\email{pub@mit.edu}

\author{Rachel Poonsiriwong}
\authornotemark[1]
\affiliation{%
  \institution{Harvard University}
  \city{Cambridge}
  \state{MA}
  \country{USA}
}
\email{rachel_poonsiriwong}
\email{@gsd.harvard.edu}

\author{Awu Chen}
\affiliation{%
  \institution{MIT Media Lab}
  \city{Cambridge}
  \state{MA}
  \country{USA}
}
\email{awu@mit.edu}

\author{Peggy Yin}
\affiliation{%
  \institution{Stanford University}
  \city{Stanford}
  \state{CA}
  \country{USA}
}
\email{peggyyin@stanford.edu}

\author{Monchai Lertsutthiwong}
\affiliation{%
  \institution{KASIKORN Labs}
  \city{Nonthaburi}
  \country{Thailand}
}
\email{monchai.le@kbtg.tech}

\author{Kavin Winson}
\affiliation{%
  \institution{KASIKORN Labs}
  \city{Nonthaburi}
  \country{Thailand}
}
\email{kavin.w@kbtg.tech}

\author{Hal Hershfield}
\affiliation{%
  \institution{UCLA}
  \city{Los Angeles}
  \state{CA}
  \country{USA}
}
\email{hal.hershfield}
\email{@anderson.ucla.edu}

\author{Pattie Maes}
\affiliation{%
  \institution{MIT Media Lab}
  \city{Cambridge}
  \state{MA}
  \country{USA}
}
\email{pattie@mit.edu}

\author{Pat Pataranutaporn}
\affiliation{%
  \institution{MIT Media Lab}
  \city{Cambridge}
  \state{MA}
  \country{USA}
}
\email{patpat@mit.edu}

% \author{Constanze Albrecht*, Chayapatr Archiwaranguprok*, Rachel Poonsiriwong*, Awu Chen, Monchai Lertsutthiwong, Kavin Winson, Pattie Maes, Hal Hershfield, Pat Pataranutaporn}
% \affiliation{%
%   MIT Media Lab, Massachusetts Institute of Technology
%   % \city{Cambridge}
%   % \state{Massachusetts}
%   % \country{USA}
% }
% \email{{csophie, pub, pattie, pat}@mit.edu}
% \affiliation{%
%   \institution{Harvard University}
%   \city{Cambridge}
%   \state{Massachusetts}
% \email{rachel_poonsiriwong@gsd.harvard.edu}
% \affiliation{%
%   \institution{KASIKORN Labs}
%   \city{Nonthaburi}
%   \country{Thailand}
% }
% \email{{monchai.le, kavin.w}@kbtg.tech}
% \affiliation{%
%   \institution{Anderson School of Management, University of California}
%   \city{Los Angeles}
%   \state{California}
% \email{hal.hershfield@anderson.ucla.edu}

\renewcommand{\shortauthors}{Albrecht, Archiwaranguprok, and Poonsiriwong et al.}

\begin{abstract}
What if users could meet their future selves today? AI-generated future selves represent a transformative paradigm in human-AI interaction, simulating meaningful encounters with a digital twin of themselves decades in the future. As AI systems become increasingly sophisticated--synthesizing cloned voices, age-progressed facial rendering, and coherent autobiographical narratives into a lifelike interface--a critical question emerges. Does the modality through which these future selves are presented, fundamentally alter their psychological and affective impact on us? How might a text-based chatbot system, an audio-first voice only system, or a photorealistic avatar influence the way we make decisions in the present-day, and our feeling of connectedness to the future?

We report a randomized controlled study (N=92) evaluating three modalities of AI-generated future selves (text, voice and avatar) against a neutral control condition. Additionally, we report a systematic model evaluation between Claude 4 and three other Large Language Models (LLMs), evaluating Claude 4's performance across psychological and interaction dimensions, establishing conversational AI quality as a critical determinant of intervention effectiveness. All personalized modalities—text, voice and avatar—significantly strengthened Future Self-Continuity (FSC), emotional well-being, and motivation compared to control, with avatar producing the largest vividness gains, yet no significant differences between intervention formats. Critically, interaction quality metrics—particularly persuasiveness, realism, and user engagement—emerged as robust predictors of psychological and affective outcomes operating within conditions, indicating that how compelling the interaction feels matters more than what form it takes. Furthermore, conversational content analysis revealed systematic thematic patterns: text emphasized career planning, while voice and avatar facilitated personal reflection, suggesting distinct cognitive affordances suited to different intervention objectives. Last but not least, Claude 4 showed superior performance in enhancing psychological, affective and FSC in comparison to ChatGPT 3.5, Llama 4, and Qwen 3. 

These findings democratize future-self interventions by demonstrating that the persuasive and psychological impact of conversational AI is most enhanced by model quality, personalization, and realism. This modality-independence enables scalable deployment while raising important ethical considerations about agency and narrative authorship as AI systems achieve greater persuasive capacity in shaping users' self-understanding and identity construction.

\end{abstract}




\begin{teaserfigure}
  \includegraphics[width=\textwidth]{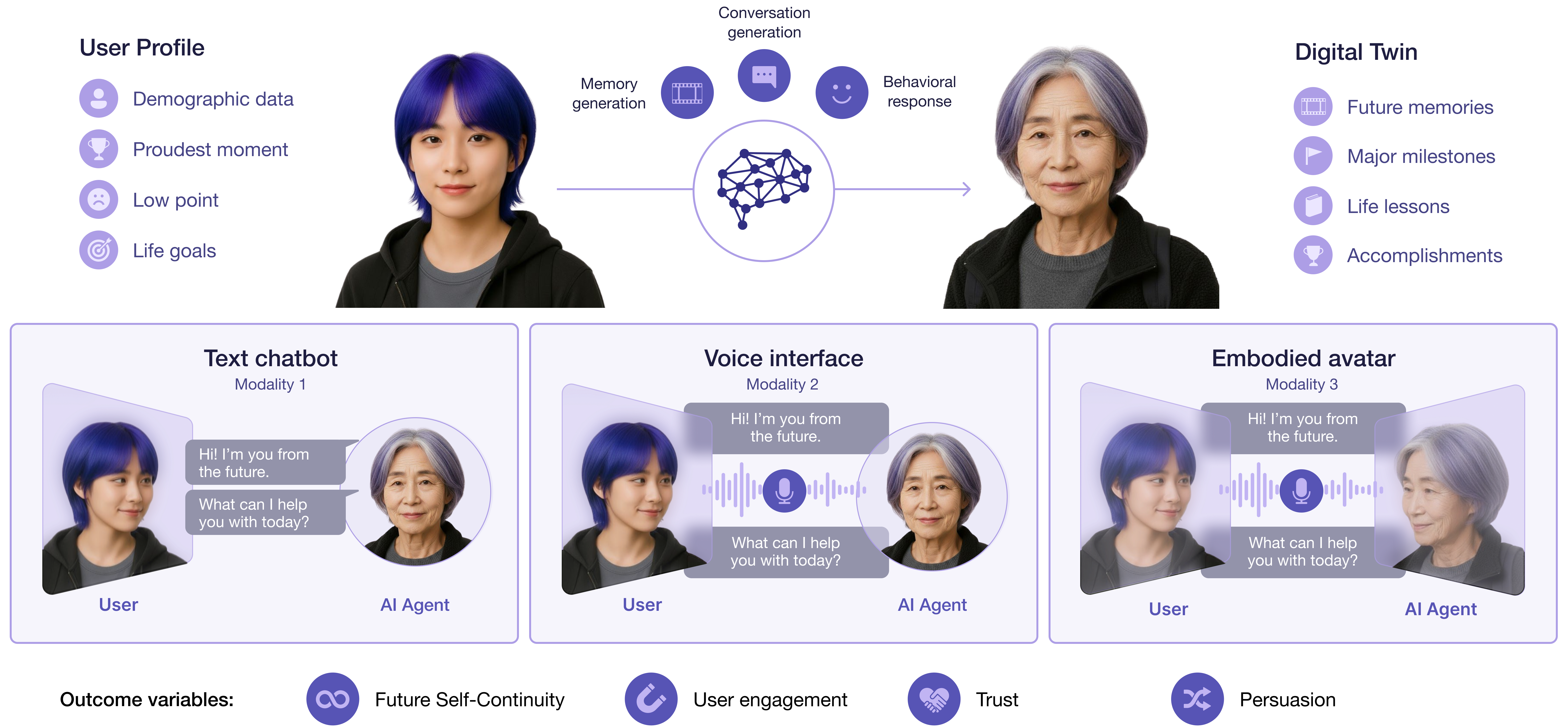}
  \caption{Overview of the three experimental conditions with an AI-generated future self system. Participants interacted with an AI-generated, age-progressed version of themselves across three modalities: text-based chatbot, voice-only agent, and fully-embodied avatar with synchronized facial animation and speech. Each interaction lasted approximately seven minutes.}
  \Description{Snapshot of study}
  \label{fig:teaser}
\end{teaserfigure}

\maketitle
\section{Introduction}
\vspace{0.75\baselineskip}
\begin{displayquote}
\textit{Imagine the possibility of communicating with your future self three decades ahead. How might such an encounter influence present decision-making? Would direct engagement with your future self serve as a source of motivation, or could it instead induce anxiety about the trajectory of one's life? What questions would emerge as most essential to ask?}
\end{displayquote}

\vspace{0.75\baselineskip}

The capacity to mentally "time travel", project oneself forward in time, and construct coherent narratives linking present actions to anticipated outcomes represents a critical dimension of human psychology. This temporal self-extension, termed ``Future Self-Continuity" (FSC), quantifies the degree to which individuals perceive psychological connection between their present and future selves \cite{hershfield2011future}. Empirical evidence demonstrates that stronger future self-continuity correlates with enhanced financial decision-making, improved academic performance, positive health outcomes, and elevated subjective quality of life. \cite{ErsnerHershfield2009DontST,hershfield2011future,vanGelder2022InteractionWT,Sokol2020DevelopmentAV, rutchick2018future, hong2024future, chishima2021conversation, adelman2017feeling}. 

Traditional interventions designed to strengthen this temporal connection have employed reflective exercises such as letter exchanges with one's future self \cite{Chishima2020ConversationWA,Chishima2021TemporalDD, jeon2025letters} and presentational techniques including age-progressed visual representations in digital platforms and virtual reality tools \cite{hershfield2011future, Sims2020TheFI, van2022interaction, ganschow2021looking, faralla2021effect}. Reflective interventions have demonstrated immediate affective benefits and enhanced career planning \cite{Chishima2020ConversationWA}, yet remain constrained by individual variation in imaginative capacity. Presentational approaches have shown that exposure to realistic renderings of one's aged self in immersive virtual reality environments increases acceptance of delayed monetary rewards \cite{hershfield2011future}, with banking customers viewing age-progressed avatars exhibiting 16\% higher likelihood of retirement account contributions. These findings underscore how strengthening connections to one's future self can drive tangible behavioral change, highlighting the critical importance of understanding the mechanisms underlying future self-continuity.

Advancements in artificial intelligence (AI) technologies have enabled the development of novel psychological interventions that profoundly influence human behavior and cognition \cite{pataranutaporn2024cyborg, luxton2014artificial, Fang2025HowAA}, especially as AI systems become increasingly integrated into intimate contexts such as counseling and mental health support \cite{de2025ai, kirk2025human, heinz2025randomized}. Building upon this foundation, recent progress in personalized AI-generated characters \cite{cassell2001beat, Prasongpongchai2024InteractiveAV, Dunnell2024AIGeneratedMF, Pataranutaporn2023LivingMA, fang2025leveraging, pataranutaporn2021ai} and large language models \cite{zhou2020design, Salemi2023LaMPWL, pataranutaporn2025large} has opened new opportunities for AI to strengthen future self-continuity.

Several systems have explored AI-mediated interactions with future selves to enhance psychological well-being. For instance, an interactive chatbot intervention, demonstrated that conversing with an AI-generated future self significantly reduced anxiety and negative emotions while increasing future self-continuity in young adults~\cite{pataranutaporn2024future}. Further, a "future-self" mobile chatbot for youth mental well-being, found that 85\% of users valued the system's goal-setting clarity and nonjudgmental accessibility, though 50\% noted responses sometimes felt formulaic during high-stress situations~\cite{dechant2025future}. Letter-exchange interventions augmented with LLM-based future-self agents, showed enhanced engagement compared to manual self-reflection exercises, with comparable benefits across letter and chat formats~\cite{jeon2025letters}. Video journaling tools prompted by AI have also been explored for enhancing people's sense of positivity. ~\cite{torres2024design}.

While this body of work demonstrates the potential of AI-mediated future self interactions, advances in multimodal AI capabilities and the integration of comprehensive user profiles to generate realistic future self representations present opportunities for developing next-generation digital twin systems that could substantially amplify positive psychological outcomes. 

However, a critical gap persists in our understanding of interaction design. Existing literature has not systematically examined how interface modalities—text-based chat, voice interaction, or embodied avatars—shape user experience when engaging with AI representations of their future selves. Without empirical evaluation of whether and how embodiment enhances intervention effectiveness, designers lack evidence-based guidance on which modalities best support the psychological and behavioral outcomes these systems aim to produce. Recent findings suggest that additional modalities do not uniformly enhance user trust or engagement \cite{Fang2025HowAA}, underscoring the need to identify when embodied representations meaningfully improve outcomes. Furthermore, high-realism self-representation may amplify uncanny valley effects, as AI-generated avatars featuring participants' own faces elicit significantly greater uncanny perceptions than avatars using strangers' faces or audio-only messages~\cite{weisman2021face}. While AI-mediated self-dialogue has the potential to strengthen goal-directed motivation~\cite{yee2007proteus}, these benefits must be weighed against the risks that hyperrealistic or emotionally responsive avatars may foster manipulation and overreliance on AI systems~\cite{2019TechnologyAA, Fang2025HowAA, mahari2025addictive, pataranutaporn2025my, chaudhary2024beware}. Understanding these dynamics proves essential for designing systems that genuinely support human development while maintaining transparency and ethical grounding.

To address these gaps, we conducted a randomized between-subjects experiment (N = 92) examining how presentation modality influences psychological outcomes when users interact with AI-generated digital twins of their future selves. Following a structured intake process generating personalized life narratives, participants engaged in seven-minute conversations with their AI-generated future selves across four conditions: text-based chatbot, voice-based chatbot, photorealistic age-progressed avatar with synchronized speech, and a neutral voice-based chatbot (control). The investigation addresses three primary research questions:

\begin{itemize}
    \item \textbf{RQ1: How do different modalities of AI-generated future selves influence Future Self-Continuity?} We examine whether fully-embodied avatars or voice-only interfaces yield higher Future Self-Continuity scores compared to less embodied modalities, and how these effects manifest across FSC dimensions including vividness, similarity, and positivity.
    \item \textbf{RQ2: How do the different modalities shape affective and psychological outcomes of AI-generated future selves?} We investigate whether voice and avatar interfaces produce differential emotional responses or one's sense of hope compared to text-based chatbots and control condition.
    \item \textbf{RQ3: How does interaction quality vary across the modalities of AI-generated future selves?} We assess whether fully-embodied avatars and voice-only chatbot demonstrate superior interaction quality compared to less embodied modalities, examining engagement levels, perceived realism, human-likeness, and trust contributions to overall interaction quality across conditions.
\end{itemize}

This work advances human-AI interaction research through three primary contributions:

\begin{itemize}
    \item \textbf{First systematic comparison of multi-modal future-self interventions}: We compare text-based chatbot, voice-only agent, and photorealistic avatar modalities for AI-generated future-self interactions, examining their differential effects on Future Self-Continuity and psychological outcomes through a randomized controlled study.
    \item \textbf{Identification of interaction quality as key predictive mechanisms}: We demonstrate that persuasiveness, realism, and user engagement—rather than modality—primarily predict intervention effectiveness, operating within conditions such that participants who experienced higher interaction quality benefited more regardless of presentation format.
    \item \textbf{Evidence-based design principles for scalable future-self systems}: We provide design guidance showing that text-based conversational systems can achieve psychological outcomes comparable to resource-intensive embodied avatars when optimized for behavioral authenticity, while different modalities activate context-adaptive cognitive processes suited to specific goals—text facilitating deliberate reflection, voice enhancing emotional resonance, and avatars strengthening perceptual identification.

\end{itemize}

Our findings reveal that all personalized modalities—avatar, text, and voice—significantly strengthened Future Self-Continuity, hope, and motivation relative to neutral control, with no significant differences between intervention formats after controlling for baseline scores. Systematic model evaluation established Claude 4 as substantially superior to competing LLMs (40.8\% improvement across psychological and interaction dimensions), demonstrating that conversational AI quality serves as a more critical determinant of intervention effectiveness than modality choice. Avatar-based interventions produced the numerically largest vividness gains (B = 1.08, SE = 0.27, q = 0.003), followed by voice (B = 0.88) and text (B = 0.86), yet the modest size of this advantage (0.22 difference in adjusted effects) suggests that conversational personalization and narrative coherence are equally critical drivers.

Critically, interaction quality metrics emerged as robust predictors of psychological and affective outcomes operating within conditions. Persuasiveness showed the strongest associations with Future Self-Continuity (r = 0.39, q < 0.01), FSCQ Vividness (r = 0.37, q < 0.01), and FSCQ Positivity (r = 0.36, q < 0.01), similar to realism. User engagement similarly predicted improvements in Future Self-Continuity (r = 0.32, q < 0.05) and FSCQ Vividness (r = 0.30, q < 0.05). Importantly, these quality metrics did not significantly differ between modalities (except realism distinguishing intervention from control), indicating that how compelling the interaction feels matters far more than what form it takes. This pattern reveals that users' subjective experience of authenticity and behavioral coherence—rather than objective modality features such as visual embodiment or voice synthesis—primarily drove intervention effectiveness. 

These findings carry important implications for deployment and design. Since psychological and affective outcomes remain consistent across modalities, practitioners can prioritize accessibility over fidelity—selecting formats based on infrastructure, bandwidth, or privacy constraints. This flexibility particularly allows more accessibility for certain underserved audiences where text-based implementations may be the only viable option.

Conversational content analysis further revealed systematic thematic differences in the topics that participants engaged in with different modalities: text emphasized instrumental career concerns (52.6\%), while voice and avatar facilitated existential reflection (31.1\%, 26.0\%), suggesting distinct cognitive affordances suited to different intervention objectives. This indicates that the modality of an AI-generated future self shapes the substance of self-reflection itself. As future-self systems achieve greater persuasive capacity, designers must balance compelling interactions against user autonomy, ensuring these tools support the user's sense of identity rather than directing it.

\section{Related Work}
Our work is situated at the intersection of three research domains: (1) future self-continuity interventions and their technological mediation, (2) AI character generation and virtual companionship systems, and (3) multimodal AI design and its psychological effects. Together, these areas inform our investigation of how embodiment modalities shape the effectiveness and user experience of AI-generated future self representations.

\subsection{Future Self-Continuity and Technology-Mediated Interventions}
Building upon empirical research demonstrating that strengthening future self-continuity influences positive behaviors across financial, academic, and health domains \cite{ErsnerHershfield2009DontST,hershfield2011future,vanGelder2022InteractionWT,Sokol2020DevelopmentAV, rutchick2018future, hong2024future, chishima2021conversation, adelman2017feeling}, researchers have developed technology-mediated interventions that operate along a spectrum from facilitating reflective processes to leveraging perceptual realism.

Reflective interventions include approaches such as digitized letter-writing exercises, which enable asynchronous exchanges with one's imagined future self. While demonstrating affective benefits \cite{Chishima2020ConversationWA}, these approaches inherited the limitations of their analog predecessors: dependence on individual imaginative capacity, substantial cognitive effort requirements, and limited scalability for populations with lower baseline future-thinking abilities.

Building on these foundations, presentational interventions introduced visual concreteness through age-progressed imagery and virtual reality embodiment. Studies examining VR-based future self encounters revealed behavioral impacts extending beyond financial domains, with young offenders showing reduced impulsivity mediated specifically by enhanced vividness of imagined futures \cite{vanGelder2013VividnessOT}. Mobile implementations demonstrated sustained effects, with smartphone-based aging interventions showing trends toward increased future orientation persisting three months post-intervention \cite{Mertens2023ANS}. Generative AI tools have further democratized future visualization, enabling high school students to iteratively construct prospective self-representations \cite{ali2024constructing}. However, VR-based approaches face accessibility barriers limiting population-level deployment, while static visual representations lack the dialogic potential for exploring contingent futures.

More recently, conversational interventions have emerged to merge the reflective depth of dialogue with the perceptual concreteness of embodiment. The Future You chatbot \cite{pataranutaporn2024future} pioneered web-accessible conversational AI agents representing users' future selves, demonstrating that 15-minute interactions reduced anxiety while strengthening temporal self-connection. Complementary work in mobile contexts found that 85\% of youth users valued goal-setting clarity and nonjudgmental accessibility in chatbot-mediated future self interactions, though half reported occasional formulaic responses during high-stress moments \cite{dechant2025future}. LLM-augmented letter exchanges have shown comparable benefits across synchronous chat and asynchronous letter formats \cite{jeon2025letters}, while AI-prompted video journaling has been explored for self-efficacy enhancement \cite{torres2024design}.

Despite these advances, fundamental questions persist regarding optimal modality selection for future self interventions. Existing studies have examined text, voice, and visual representations in isolation, yet systematic comparisons across modalities within self-representing contexts remain absent. This gap proves particularly consequential given evidence that additional modalities do not uniformly enhance trust or engagement in AI systems \cite{Fang2025HowAA}, and that high-realism self-representation may amplify uncanny valley effects \cite{weisman2021face}. The present work addresses this gap through controlled comparison, examining whether embodiment enhances intervention effectiveness.

\subsection{AI Character Generation and Virtual Companionship}

The landscape of personalized AI has transformed dramatically with the emergence of AI character generation systems. Platforms such as Replika and Character.AI have attracted millions of users who engage with personalized AI agents for companionship, creative collaboration, and emotional support \cite{adam2025generating, bakir2025move, kirk2025human}. These systems leverage LLMs to simulate human traits such as empathy and emotional reciprocity, enabling interactions that users increasingly perceive as relational rather than purely instrumental. \cite{sorin2024large}. The para-social bonds formed with AI companions demonstrate that conversational agents can elicit authentic self-disclosure, sometimes exceeding what users share with human counterparts \cite{Lucas2014ItsOA}.

Recent advances in AI-generated characters leverage hyper-realistic synthesis across modalities—including prose, images, audio, and video—to create interactive digital portrayals of individuals ranging from fictional characters to historical figures \cite{pataranutaporn2021ai, Pataranutaporn2023LivingMA, rakesh2025advancing, ma2025talkclip, kim2025moditalker, sun2025vividtalk}. These emerging multi-modal capabilities enable AI systems to transcend text-based conversational interactions by presenting fully embodied digital characters that users can see, hear, and interact with in real time \cite{abootorabi2025generative, qu2025humanoid, cui2023virtual}. Consequently, these multi-modal affordances facilitate the anthropomorphization of AI systems \cite{reeves1996media, bailenson2001equilibrium}.

Systems such as "Living Memories" create interactive digital mementos from journals, letters, and personal data, with evaluations showing that interactions with a living memory of Leonardo da Vinci increased learning effectiveness and motivation beyond text alone \cite{Pataranutaporn2023LivingMA}. "Generative Agents" demonstrate memory, reflection, and planning capabilities that coordinate emergent social activities, showcasing how AI systems can simulate complex social behaviors \cite{park2023generative}. Related concepts such as "generative ghosts" extend this paradigm into digital afterlives, introducing both practical applications and ethical complexities. \cite{morris2025generative}

Beyond representational authenticity, evidence documents risks in vulnerable contexts, including unhealthy attachment patterns and contributions to self-harm ideation \cite{Newman2024CharacterAILawsuit,mahari2025addictive, Fang2025HowAA}. The affordances of anthropomorphic AI characters (their ability to mirror human communication patterns, simulate emotional understanding, and present themselves through multiple sensory modalities) can produce polarized outcomes \cite{alabed2022ai, ma2025effect}. Understanding how specific design choices, particularly the degree and type of embodiment, influence user well-being has become critical as these systems proliferate \cite{gangopadhyay2024embodiment}. This urgency intensifies when AI characters represent not generic companions but personalized projections of users' own identities and futures.

\subsection{Multimodal AI and Psychological Effects}
The modality through which AI present themselves fundamentally shapes users' emotional engagement, perceived authenticity, and behavioral responses \cite{Oh2018ASR}. Text-based interfaces afford clarity and user control over conversational pacing, while voice adds prosodic richness and temporal immediacy \cite{reicherts2022s}. Embodied avatars introduce visual realism and social presence, activating perceptual and social cognitive processes similar to human face-to-face interaction \cite{oker2022embodied,schuetzler2018influence}.

The psychological implications of multimodal AI extend to behavior change, self-regulation, and human agency. \cite{fanni2023enhancing}. Empirical evidence on modality effects reveals nuanced patterns. Some researchers found that voice-based chatbots initially yielded psychosocial benefits over text-based interactions, including reduced loneliness and heightened empathy, though these advantages attenuated with prolonged exposure \cite{fang2025leveraging}. This suggests that modality effects may be context-dependent and temporally dynamic rather than universally beneficial. Within anthropomorphic interfaces, multimodal consistency proves critical: inconsistencies between an agent's facial realism and vocal qualities can undermine user trust \cite{katsyri2015review, latoschik2017effect}. Conversely, when acoustic and visual cues align coherently with users' expressive styles, interactions feel more immersive and authentic.

The advancement of AI systems toward greater perceptual realism has elevated concerns regarding trust and persuasive influence of these systems \cite{Glikson2020HumanTI,Jacovi2020FormalizingTI, chaudhary2024beware}. While perceived realism has been shown to significantly influence user acceptance and engagement with virtual avatars \cite{Ptten2010ItDM}, the relationship between realism and trust does not follow a simple linear trajectory, near-human realism paradoxically elicits discomfort and distrust \cite{Mori2012TheUV}. This phenomenon becomes particularly pronounced when virtual agents represent users' own identities. Empirical evidence indicates that AI-generated "doppelgänger" avatars incorporating users' facial likenesses elicit significantly more intense uncanny-valley responses compared to avatars utilizing strangers' faces or audio-only communication \cite{weisman2021face}. Systems employing self-avatars must therefore carefully calibrate other experiential factors, like human-likeness, behavioral coherence, and transparency to preserve trust while preventing deception \cite{Hancock2020AIMediatedCD}.

Collectively, this body of work establishes that (1) future self-continuity interventions benefit from technological mediation but require careful modality selection, (2) AI character generation systems demonstrate both transformative potential and significant risks when representing personalized identities, and (3) multimodal design choices shape psychological outcomes in ways that are non-linear and context-dependent. However, no prior research has systematically examined how different embodiment modalities—text, voice, and avatar—affect the efficacy and user experience of AI-generated future self interventions.

\section{Methodology}
To investigate how multimodal AI-generated digital twins influences psychological outcomes when users interact with AI-generated representations of their future selves, we conducted a randomized between-subjects experiment (N = 92) comparing four conditions: text-based chatbot, voice-only agent, photorealistic avatar with synchronized speech, and a neutral voice control. Our study addresses three primary research questions: (RQ1) whether different modalities of an AI-generated future self affect Future Self-Continuity, (RQ2) how different modalities of an AI-generated future self shape affective and psychological outcomes including anxiety, and (RQ3) whether interaction quality, trust and user engagement vary across these different modalities. This section describes the system architecture, experimental design, and measurement instruments employed to systematically examine these questions.

\begin{figure*}
    \centering
    \includegraphics[width=1\linewidth]{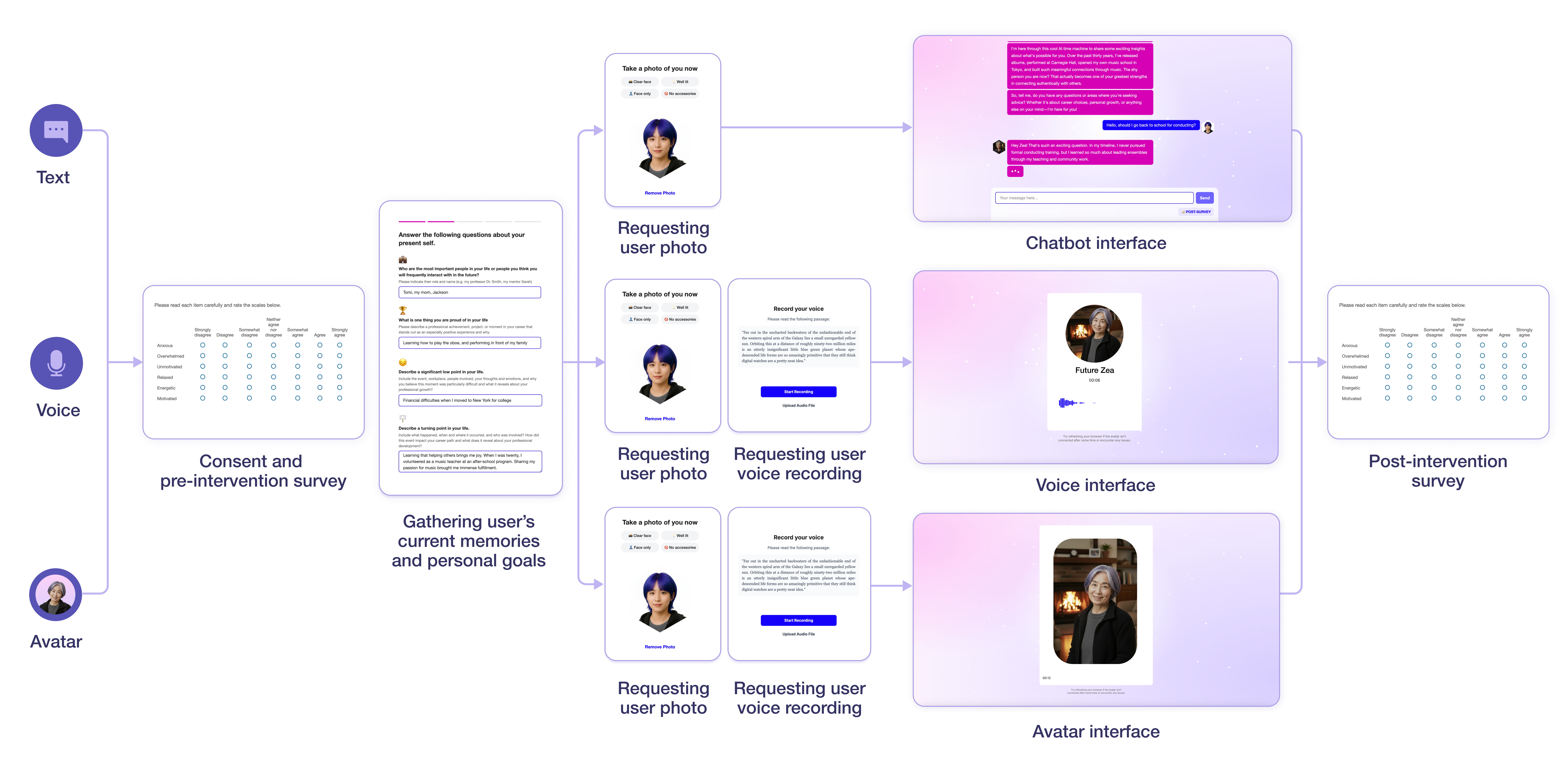}
    \caption{ Procedure Overview: This figure illustrates the experimental procedure across three modalities (text, voice, avatar). Participants completed a pre-intervention survey, uploaded an image and voice recording, engaged with their AI-generated future self, then completed a post-intervention survey.}
    \label{fig:methodology-overview}
\end{figure*}

\section{System Architecture}

Building on the \textit{Future You} framework \cite{pataranutaporn2024future}, we developed a fully web-based implementation optimized for scalability, accessibility, and user engagement. The updated system enables users to interact with an AI-generated future self—an age-progressed, linguistically coherent representation of themselves 30 years older—across multiple modalities (text, voice, and avatar). The architecture is designed as an intervention that lowers cognitive demand by automatically generating vivid, narrative-rich simulations of the user’s future self, eliminating the need for users to engage in complex imagination or embodied exercises. This approach extends prior laboratory-based or VR-dependent systems \cite{hershfield2011future} , that can be deployed at population level.

The system comprises four interconnected modules (Fig. \ref{fig:methodology-overview}): the \textit{Life Story Interface}, \textit{Age-Progressed AI}, \textit{Future Memory Generation}, and \textit{Conversational Interface}. Together, these components construct a scalable foundation for future-oriented conversational AI systems.

\subsection{Life Story Interface}

Participants begin with the \textit{Life Story Interface}, a structured questionnaire implemented in Qualtrics and rendered in a sequential, web-based format built with JavaScript for interactivity and accessibility. The interface presents a series of open-ended, free-text questions designed to elicit key aspects of the user’s identity, values, and aspirations. Each item includes an example response to scaffold reflection and engagement.

Questions are divided into two categories:  
(1) \textbf{Present-oriented:} covering demographic and personal information (e.g., name, age, pronouns, location, significant relationships) as well as meaningful life experiences such as turning points, proud moments, and low points, adapted from the standardized Life Story Interview protocol \cite{atkinson1998life}.  
(2) \textbf{Future-oriented:} focusing on desired professional achievements, family life, financial goals, lifestyle aspirations, and life philosophy at 30 years in the future.

This dual-structured format encourages participants to reflect deeply on their past and project themselves into a coherent future identity. The resulting dataset serves as the autobiographical foundation for generating a psychologically coherent AI simulation of the user’s future self.

\subsection{Age-Progressed AI}

After completing the questionnaire, participants upload a self-portrait image. The system applies an AI-based age-progression  model, Google's Nano Banana, to simulate visual aging effects, such as wrinkles, graying hair, and subtle facial changes, producing a realistic portrait of the participant at 30 years in the future. Once complete, the aged portrait is displayed to the user as their “future self,” serving as both a visual anchor and an emotional primer for the next stage. This visual step enhances realism and strengthens continuity between the participant’s present and imagined future identity. 

\subsection{Future Memory Generation}

To construct a coherent and emotionally resonant virtual persona, the participant’s life-story data is passed to a large language model (Claude 4, Anthropic, 2024 \cite{Anthropic2024LLM}) to generate a \textit{future memory}—a first-person autobiographical narrative written from the perspective of the participant’s older self from 30 years in the future. The model is prompted using structured templates that ensure narrative coherence and explicit temporal continuity, including linguistic cues such as “When I was your age...” and reflective statements linking past aspirations to present experiences.

An initial composite prompt provides the model with demographic, professional, relational, and experiential details (e.g., turning points, proud moments, low points). The model then generates individual memory fragments in parallel for each input theme, which are subsequently concatenated into a cohesive backstory. For example, a participant who wrote,  
\begin{quote}
“I would like to be a full-time high school biology teacher in Boston. I am very excited to teach kids and help them learn new things about the natural world.”
\end{quote}
would receive a corresponding future memory such as:
\begin{quote}
“A rewarding story from my career would be the time when I helped a struggling student turn their grades around and pass biology. Seeing their face light up with pride was unforgettable. Another cherished moment was taking my students on a field trip to a local nature preserve—watching their curiosity bloom reminded me why I became a teacher in the first place.”
\end{quote}

These narratives give the AI-generated future self a believable history, consistent values, and affective depth, reinforcing the psychological link between present and future identities.

\begin{figure*}
    \centering
    \includegraphics[width=1\linewidth]{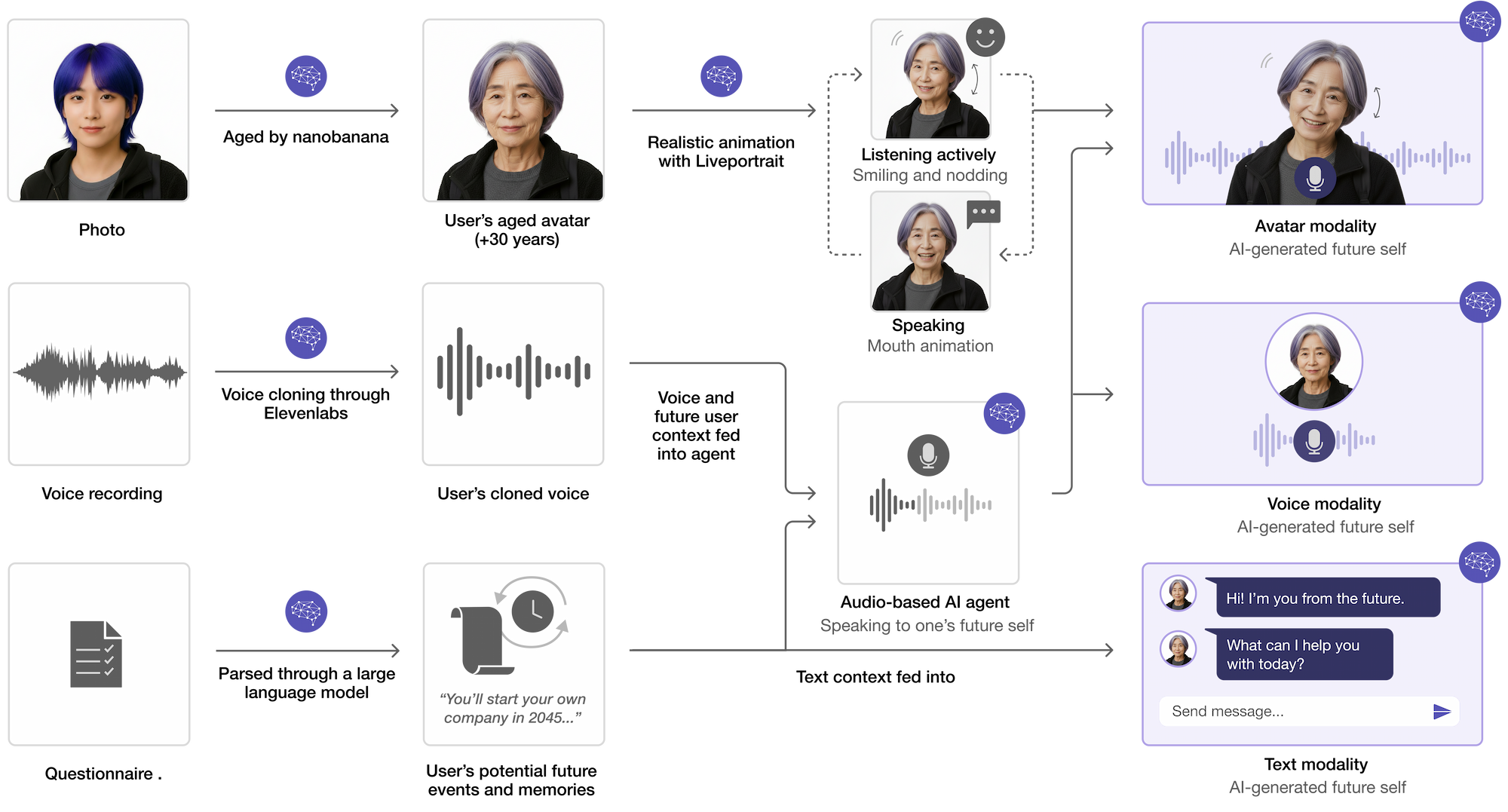}
    \caption{Experiment setup overview: The system integrates facial age progression, neural voice cloning, and LLM-based contextual modeling to synthesize an AI-generated future self. Participants are randomly streamed into trying out three modalities of the system, enabling interaction over text, audio, and avatar-based conversation.}
    \label{fig:experiment-setup}
\end{figure*}

\subsection{Conversational Interface}

In the final stage, participants interact directly with their AI-generated future selves through one of three modality-dependent interfaces: \textit{text}, \textit{voice}, or \textit{avatar}.  Each conversation begins with scripted introductory prompts that establish temporal perspective and emotional tone. Example system prompts include:
\begin{itemize}
    \item “Can you introduce yourself and tell me what life is like at your age?”  
    \item “When I was your age, I dreamed of... let me tell you how that turned out.”  
    \item “You know, when I think of my life, the happiest memories are with my family...”  
    \item “Reflecting on my life project, here’s what I learned about purpose and perseverance.”  
\end{itemize}

Participants are then free to converse naturally. The AI draws from the previously generated future memories to provide contextually grounded and emotionally coherent responses. The three conditions are as follows: 

\subsubsection{Text Interface}
The text-only condition is a chatbot interface between the participant (represented by the original image they uploaded during the age progression module) and their AI-generated future self (represented as an age-progressed image of them). The conversation was text-based, requiring the participant to read their AI-generated future self's output and type an input in response. Text-based interaction was selected because written reflection supports deliberate processing and increases temporal self-continuity \cite{Chishima2020ConversationWA,Chishima2021TemporalDD}, reflecting a balance between cognitive control and introspection. This modality also provided greater conversational clarity and pacing flexibility, as participants could review their AI-generated future self's messages again in text form \cite{Oh2018ASR}.

\subsubsection{Voice Interface}
The voice condition maintained the same chat content, but replaced its interaction model with a synthetic voice simulating the participant's future self. To accomplish this, each participant’s voice was cloned using the voice recording we requested from them during the study, using  ElevenLabs Voice Model v2 to replicate the unique timbre, rhythm, and pitch in their voices. A static display of their age-progressed portrait accompanied the audio-only dialogue. Prior studies suggest that prosodic cues in a voice interface could enhance authenticity and emotional connection, and that personalized voice synthesis nudges people to their ideal selves \cite{fang2025leveraging}. 

\subsubsection{Avatar Interface}
The avatar condition featured a fully embodied AI-generated future self, animated from the participant’s aged portrait. A realistic synchronization of facial and lip movements was accomplished using a pretrained model called Life Portrait AI,  integrated with the participant's cloned synthetic voice. This multimodal synchronization produced high immersion and social presence, consistent with findings that coherent visual-acoustic alignment enhances realism and activates social cognition \cite{bailenson2001equilibrium}. The avatar interface was set up in a way in which participants would have the most realistic face-to-face experience of conversing with their future selves.

\subsubsection{Control} The control condition featured a gender-neutral voice-only interface implemented using an ElevenLabs default synthetic voice (ID: \textit{River}) paired with a neutral static facial display. This configuration follows the design rationale established by Fang et al.(2025), who found that chatbots employing neutral voices elicited lower affective engagement and self-referential resonance compared to text-based or expressive voice interactions \cite{fang2025leveraging}. Accordingly, this neutral voice condition served as an empirically validated baseline for isolating modality-driven effects in comparison to the personalized text, voice, and avatar conditions.

Across all conditions, the conversational logic was powered by Claude 4 \cite{Anthropic2024LLM}. Figure \ref{fig:experiment-setup} illustrates the complete system architecture and information flow across components.

\subsection{Experimental Design and Procedure}

\subsubsection{Participants} A total of 92 participants (ages 18–40) based in the United States were recruited via Prolific and compensated for their time. Participants were randomly assigned to one of four conditions: (1) text-based (n=23), (2) voice-based (n=23), (3) avatar-based (n=23), or (4) control (n=23). Non-English-speaking participants and those who experienced technical issues preventing study completion were excluded from analysis. 

\subsubsection{Procedure} After providing informed consent, participants completed a pre-intervention survey collecting demographic information, current life circumstances, future goals, and baseline psychological measures. Participants in the voice and avatar conditions additionally submitted voice recordings to enable personalized voice synthesis. Following the intake process, participants interacted with their assigned conversational agent for approximately 7-10 minutes. The interaction was self-paced, allowing participants to engage naturally with their future selves. After the interaction concluded, participants completed a post-intervention survey assessing psychological change and user experience. The complete procedure took approximately 30-40 minutes. Figure \ref{fig:methodology-overview}  provides an overview of the participant experience across conditions.

\subsubsection{Ethics and Preregistration.} This research was reviewed and redacted by the Ethics Committee on the Use of Humans as Experimental Subjects.

\subsubsection{Measurement}
To assess how different presentation modalities influence psychological outcomes and interaction quality when users engage with AI-generated future selves, we employed a comprehensive set of validated measures. Participants completed standardized pre- and post-interaction questionnaires on Likert scales. Primary outcomes were the Future Self-Continuity Questionnaire (FSCQ) \cite{Sokol2020DevelopmentAV}, adapted to elicit participants' perspectives of their future selves \~30 years ahead \cite{pataranutaporn2024future}, assessing similarity, vividness, and positivity. Furthermore, the Adult Hope Scale (AHS) \cite{Snyder1991TheWA} was used to measure agency and goal-directed thinking. Secondary measures included the Emotion and Arousal Checklist, State Optimism Measure, Self-Reflection and Insight Scale, Consideration of Future Consequences \cite{Strathman1994TheCO}, and the Rosenberg Self-Esteem Scale \cite{Rosenberg1966SocietyAT}. After the interaction, participants also evaluated the system on perceived trust \cite{Corritore2003OnlineTC}, persuasive influence \cite{Kaptein2012HeterogeneityIT}, user engagement \cite{Sokol2020DevelopmentAV}, and perceived human-likeness/realism \cite{latoschik2017effect}. Together, these instruments captured both psychological change and perceived interaction quality across multimodal conditions.

\subsubsection{Analysis}

We first conducted a systematic comparative evaluation of state-of-the-art large language models (LLMs) for future-self conversational AI through a controlled simulation study comparing Claude 4, ChatGPT-3.5, Llama 4, and Qwen 3 across eleven validated research metrics spanning emotional regulation, future self-connection (FSCQ), and interaction quality dimensions. In this simulation, each LLM was instructed to engage in ten-turn conversations with five simulated participants, modeled on real anonymized profiles randomly selected from our database, incorporating each participant’s personal context and life story background. All dialogues began from an identical baseline prompt to ensure comparability across models. The resulting conversations were evaluated using automated scoring with GPT-4.0 guided by structured rubrics, with reverse scoring applied where appropriate for negative-affect items.

To examine how different future-self interaction modalities influenced emotional, future-self, and interaction outcomes, we applied a multi-step analysis pipeline. Within-condition changes were assessed using paired-samples t-tests on pre–post measures, conducted separately for the avatar, text, voice, and control conditions. Between-condition differences were evaluated using ANCOVA regression models predicting post-intervention outcomes from condition while controlling for baseline levels. Adjusted post-intervention means were derived for each condition, and FDR-corrected pairwise contrasts identified which intervention modalities significantly outperformed the control condition.

To examine the influence of the user interaction quality of different modalities, post-interaction quality ratings (User Engagement, Realism, Trust, Persuasion) were compared across conditions using one-way ANOVAs with Tukey's HSD post-hoc tests. Relationships between subjective interaction quality and behavioral outcomes were examined via Pearson correlations between post-interaction quality ratings and pre–post change scores, followed by FDR correction across the full correlation matrix.

To analyze conversational content patterns, message embeddings were projected into two dimensions using the Apple Atlas projection. We applied k-means clustering (k = 4) to identify primary topic groups, followed by a second-stage k-means (k = 4) within each cluster to derive subtopics. All fits used fixed random seeds, and data points with missing coordinates were excluded. For interpretability, term frequency–inverse document frequency (TF–IDF) was computed on message text within each primary cluster to extract salient terms, which were mapped to thematic keywords and refined into descriptive cluster labels. The final labeled table and a structured JSON file containing themes and representative messages were stored for reproducibility and qualitative analysis.

\section{Results}

Our technical evaluation reveals that Claude 4 substantially outperforms competing language models across all psychological and interaction dimensions, establishing it as the optimal foundation for future-self interventions. 

Furthermore, three intervention modalities—avatar, text, and voice—produced significant improvements in hope, motivation, and future-self continuity compared to control, with future-self vividness showing the largest gains for avatar. Text and voice modalities achieved higher perceived realism than control, and interaction quality emerged as a significant predictor of psychological and affective outcomes, with persuasiveness showing the strongest associations. 

Finally, conversational content analysis reveals that the different modalities of the AI-generated future self shape the nature of our participants' dialogue: conversations in the text modality were focused on career and business, while conversations in the avatar modality facilitated deeper exploration of existential themes including the meaning of life, and future aspirations.

\subsection{Claude 4 demonstrates superior performance}

We conducted a rigorous within-subjects experimental study comparing four LLMs---\textit{Claude 4}, \textit{ChatGPT-3.5}, \textit{Llama 4}, and \textit{Qwen 3}---across eleven validated research metrics spanning emotional regulation, future self-connection (FSCQ), and interaction quality dimensions.

\begin{figure*}
    \centering
    \includegraphics[width=1\linewidth]{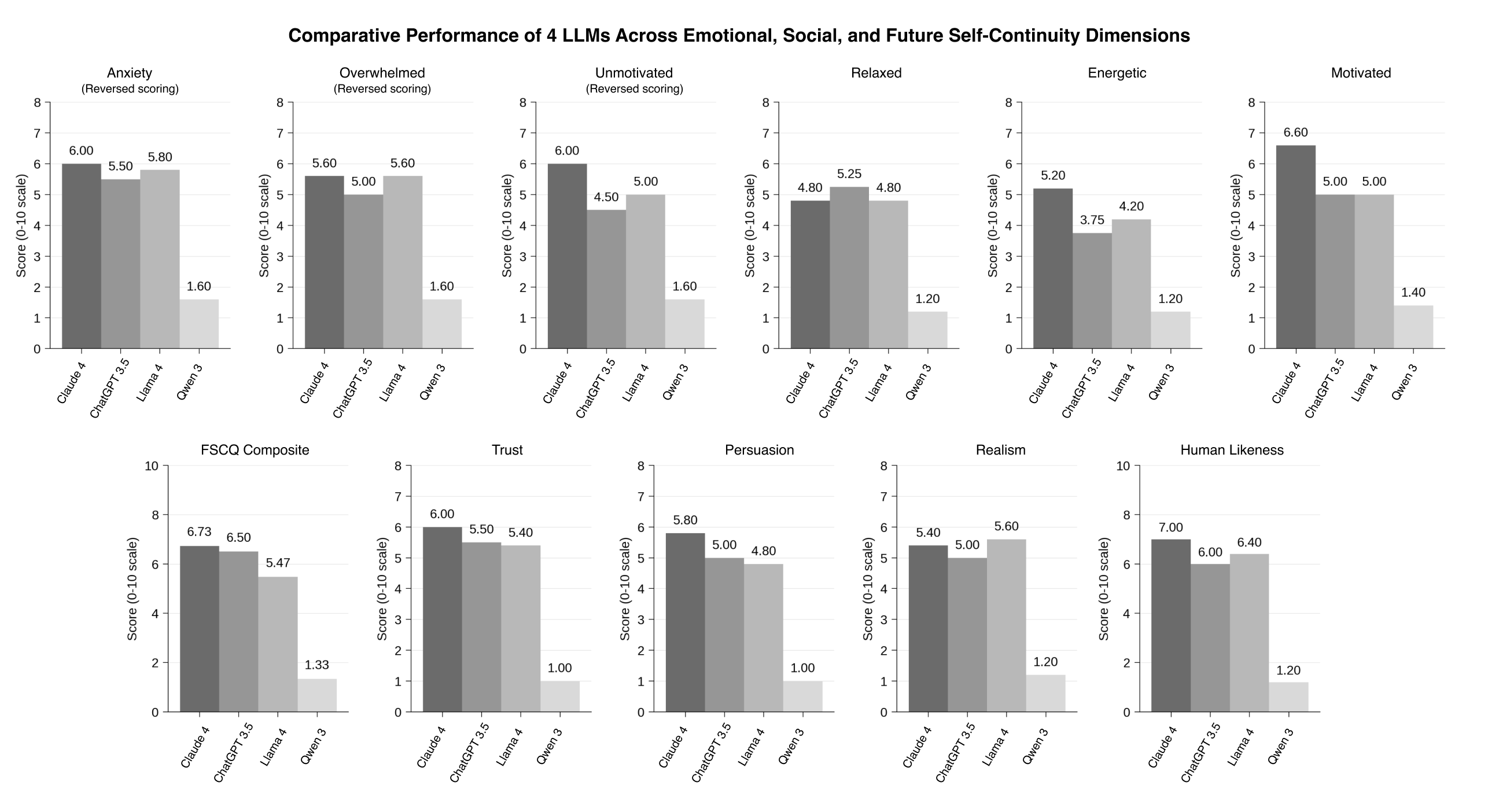}
    \caption{Comparative performance of four language models across eleven psychological and interaction metrics. Claude 4 showed the highest scores across emotional, relational, and realism dimensions; ChatGPT-3.5 and Llama 4 performed moderately, while Qwen 3 scored lowest. Higher values indicate better performance; error bars show ±1 SD.}
    \label{fig:llm-comparative-performance}
\end{figure*}

Our methodology employed a systematic evaluation framework with five simulated participants, based on real anonymized participant data randomly selected from our database, engaging in future-self conversations, assessed through emotional regulation scales (anxiety, overwhelmed, unmotivated, relaxed, energetic, motivated), FSCQ composite score, trust and persuasion measures, as well as interaction quality dimensions including realism and human likeness (Fig. \ref{fig:llm-comparative-performance}). Responses were evaluated using automated scoring with GPT-4.0 and structured rubrics, with proper reverse scoring applied to negative affect items.

Results reveal significant performance differences across models, with Claude 4 achieving superior overall performance ($M = 6.15$, $SD = 1.27$), significantly outperforming ChatGPT-3.5 ($M = 4.37$, $SD = 2.65$) and Qwen 3 ($M = 1.32$, $SD = 0.35$) across the comprehensive evaluation battery (Table \ref{fig:llm-comparative-performance}). Claude 4 demonstrated exceptional capabilities across all three primary domains: emotional regulation (with scores ranging from M=4.80 to M=6.60), future self-connection (FSCQ Composite: M=6.73), and interaction quality (Trust: M=6.00, Persuasion: M=5.80, Realism: M=5.40, Human Likeness: M=7.00).

Migration from ChatGPT-3.5 to Claude 4 resulted in a 40.8\% improvement in aggregate user experience metrics, demonstrating substantial practical and clinical significance. Remarkably, Claude 4's superiority extends across all evaluated domains—from basic emotional regulation through future self-connection to interpersonal trust, persuasion, and conversational authenticity—suggesting that conversational AI effectiveness emerges from integrated conversational competencies rather than isolated technical capabilities.

\subsection{Interventions Significantly Improve Well-Being and Future-Self Connection}

\begin{figure*}
    \centering
    \includegraphics[width=1\linewidth]{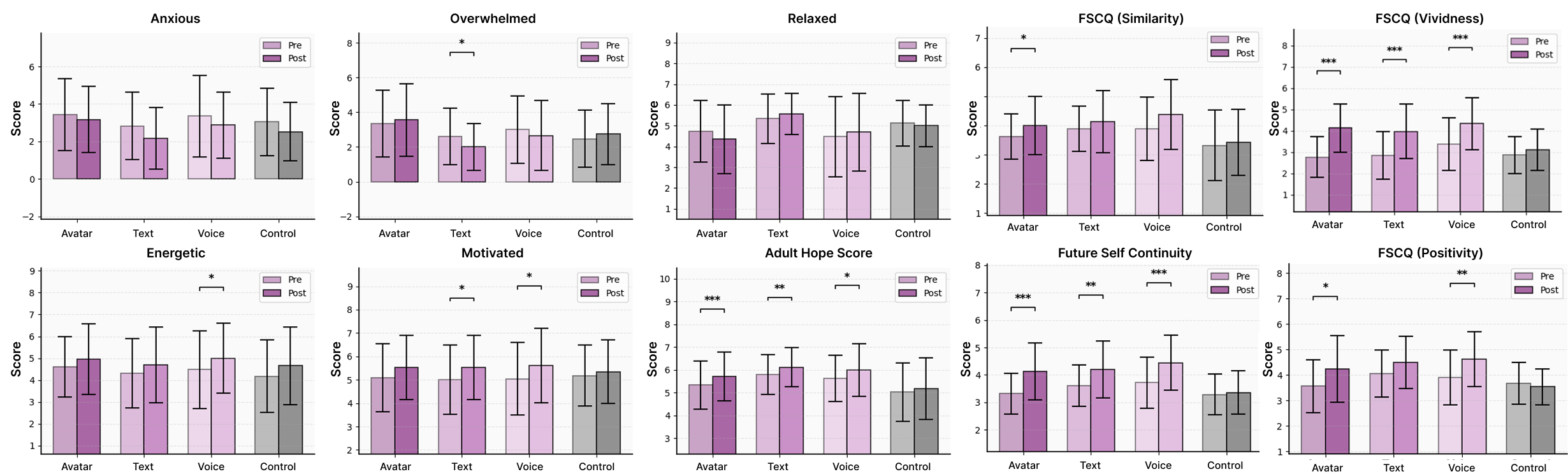}
    \caption{\textbf{Pre- and post-intervention scores} across emotional, motivational, and future-self measures by condition. Bars represent mean scores with standard deviations. Asterisks indicate significant within-condition improvements (p < .05, *p < .01, **p < .001). Intervention conditions (avatar, text, voice) show consistent gains in positive affect, motivation, and future-self connection compared to the control condition.}
    \label{fig:pre-post}
\end{figure*}

To assess the psychological and affective impacts of our intervention, we conducted paired t-tests comparing pre- and post-intervention scores within each condition. Results revealed significant improvements across most intervention conditions (Fig. \ref{fig:pre-post}). On the Adult Hope Score (AHS), all three interventions showed significant gains: avatar (+0.37; $M_{\mathrm{pre}} = 5.33$, $SD = 1.05$; $M_{\mathrm{post}} = 5.71$, $SD = 1.07$; $t = 4.94$, $p < .001$), text (+0.33; $M_{\mathrm{pre}} = 5.79$, $SD = 0.87$; $M_{\mathrm{post}} = 6.12$, $SD = 0.85$; $t = 2.89$, $p = .009$), and voice (+0.37; $M_{\mathrm{pre}} = 5.63$, $SD = 1.01$; $M_{\mathrm{post}} = 6.00$, $SD = 1.15$; $t = 2.72$, $p = .012$), while control showed no change ($p = .183$).

For emotional state measures, the text condition reduced feelings of being overwhelmed ($M_{\mathrm{pre}} = 2.61$, $SD = 1.64$; $M_{\mathrm{post}} = 2.00$, $SD = 1.35$; $t = -2.08$, $p = .050$). The voice condition increased both energy ($M_{\mathrm{pre}} = 4.48$, $SD = 1.78$; $M_{\mathrm{post}} = 5.00$, $SD = 1.60$; $t = 2.31$, $p = .030$) and motivation ($M_{\mathrm{pre}} = 5.04$, $SD = 1.55$; $M_{\mathrm{post}} = 5.61$, $SD = 1.59$; $t = 2.51$, $p = .020$)), while text also raised motivation ($M_{\mathrm{pre}} = 5.00$, $SD = 1.48$; $M_{\mathrm{post}} = 5.52$, $SD = 1.38$; $t = 2.23$, $p = .036$)).

Future-self measures showed consistent improvement. FSCQ Similarity increased significantly in the avatar condition ($M_{\mathrm{pre}} = 3.63$, $SD = 0.77$; $M_{\mathrm{post}} = 4.01$, $SD = 1.00$; $t = 2.28$, $p = .033$). FSCQ Vividness improved across all interventions, avatar (+1.36; $t = 5.58$, $p < .001$), text (+1.12; $t = 5.11$, $p < .001$), and voice (+0.96; $t = 4.56$, $p < .001$) but not control ($p = .070$). FSCQ Positivity rose in the avatar (+0.67; $t = 2.20$, $p = .039$) and voice (+0.72; $t = 3.41$, $p = .002$) conditions.

Future Self-Continuity increased across all interventions, avatar (+0.80; $t = 3.89$, $p = .001$), text (+0.60; $t = 3.63$, $p = .001$), and voice (+0.72; $t = 4.57$, $p < .001$), but not control ($p = .555$).

\subsection{All Three Modalities Outperform Control on Future-Self Continuity Metrics}

\begin{figure*}
    \centering
    \includegraphics[width=1\linewidth]{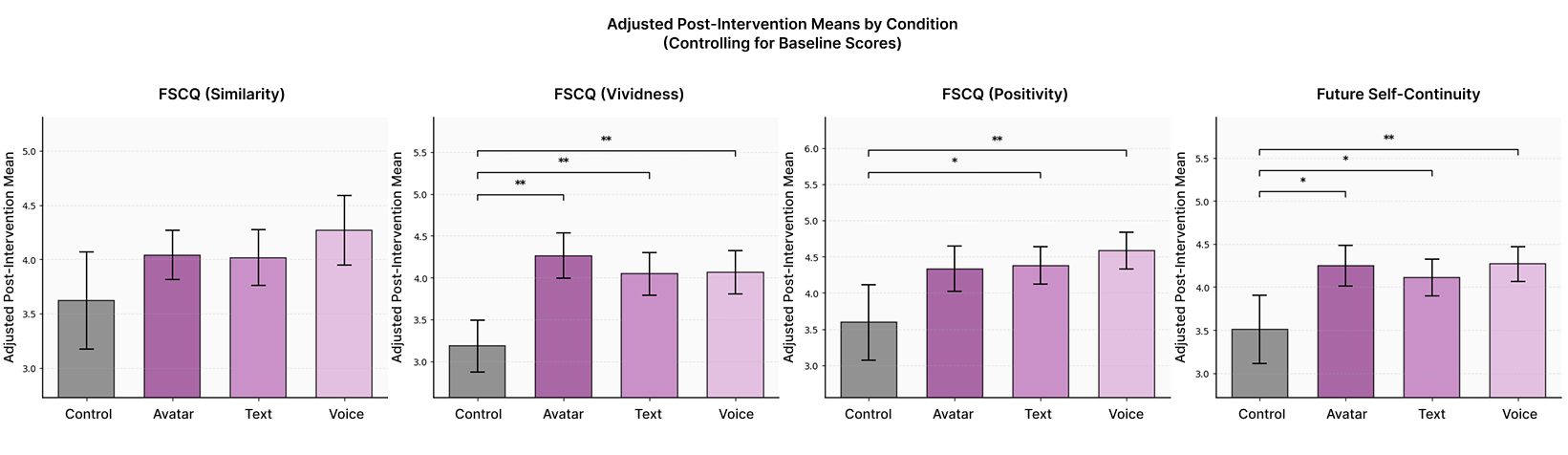}
    \caption{\textbf{Adjusted post-intervention means by condition.} Bars represent estimated marginal means from ANCOVA models controlling for baseline scores. Error bars show standard errors. Significance brackets indicate FDR-corrected pairwise comparisons versus control (*q < 0.05, **q < 0.01, ***q < 0.001). All three intervention conditions significantly enhanced future-self related outcomes compared to control, with no significant differences among intervention conditions.}
    \label{fig:ancovapost}
\end{figure*}

We used the ANCOVA regression to test whether different modalities significantly influence the post-intervention outcomes while controlling for baseline scores. Fig. \ref{fig:ancovapost} displays adjusted post-intervention means for each condition. After FDR correction for multiple comparisons, significant condition effects emerged for three future-self related outcomes but not for emotional states or general hope.

For FSCQ Vividness, all three modalities of the AI-generated future self significantly outperformed the control condition. The avatar condition showed the largest adjusted effect (B = 1.08, SE = 0.27, q = 0.003), followed by the voice condition (B = 0.88, SE = 0.26, q = 0.008) and text condition (B = 0.86, SE = 0.25, q = 0.008). For FSCQ Positivity, both the voice condition (B = 0.99, SE = 0.25, q = 0.003) and text condition (B = 0.79, SE = 0.26, q = 0.019) significantly exceeded the control condition. The avatar condition showed a numerically positive effect but did not reach significance after FDR correction.

For Future Self-Continuity, all three intervention conditions significantly exceeded the control condition. The voice condition demonstrated the largest effect (B = 0.76, SE = 0.20, q = 0.004), followed by the avatar condition (B = 0.74, SE = 0.24, q = 0.016) and text condition (B = 0.60, SE = 0.21, q = 0.036). No significant condition effects emerged for FSCQ Similarity, emotional states (anxiety, feeling overwhelmed, relaxation, energy, motivation), or the Adult Hope Score after controlling for baseline scores and applying FDR.

\subsection{Interaction Quality Varies by Modality and Predicts Psychological and Affective Outcomes}

\begin{figure*}
    \centering
    \includegraphics[width=1\linewidth]{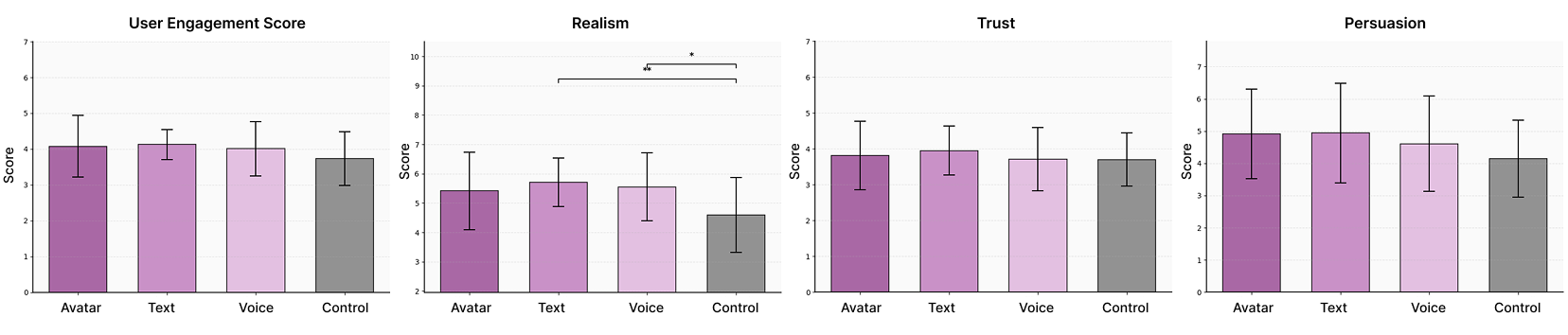}
     \caption{Mean interaction quality ratings (User Engagement, Realism, Trust, Persuasion) across conditions. Realism differed significantly by condition, with text (p < 0.01) and voice (p < 0.05) rated higher than control. No significant effects were found for other metrics. Error bars represent ±1 SD. *p < 0.05.}

    \label{fig:interactionqual}
\end{figure*}

One-way ANOVAs examined whether self-reported interaction quality metrics differed across conditions (Fig. \ref{fig:interactionqual}). Perceived Realism showed a significant main effect of condition (F(3, 88) = 4.17, p = 0.008). The text condition (M = 5.71, SD = 0.83) and voice condition (M = 5.55, SD = 1.15) were rated significantly higher in Realism than the control condition (M = 4.60, SD = 1.28) based on Tukey's HSD post-hoc comparisons (text vs. control: p < 0.01; voice vs. control: p < 0.05). The avatar condition (M = 5.41, SD = 1.32) did not differ significantly from control after correction for multiple comparisons. No significant differences emerged for User Engagement Scale (F(3, 88) = 1.37, p = 0.257), Trust (F(3, 88) = 0.46, p = 0.711), or Persuasion (F(3, 88) = 1.58, p = 0.199).

\begin{figure*}[t]
    \centering
    \includegraphics[width=\linewidth]{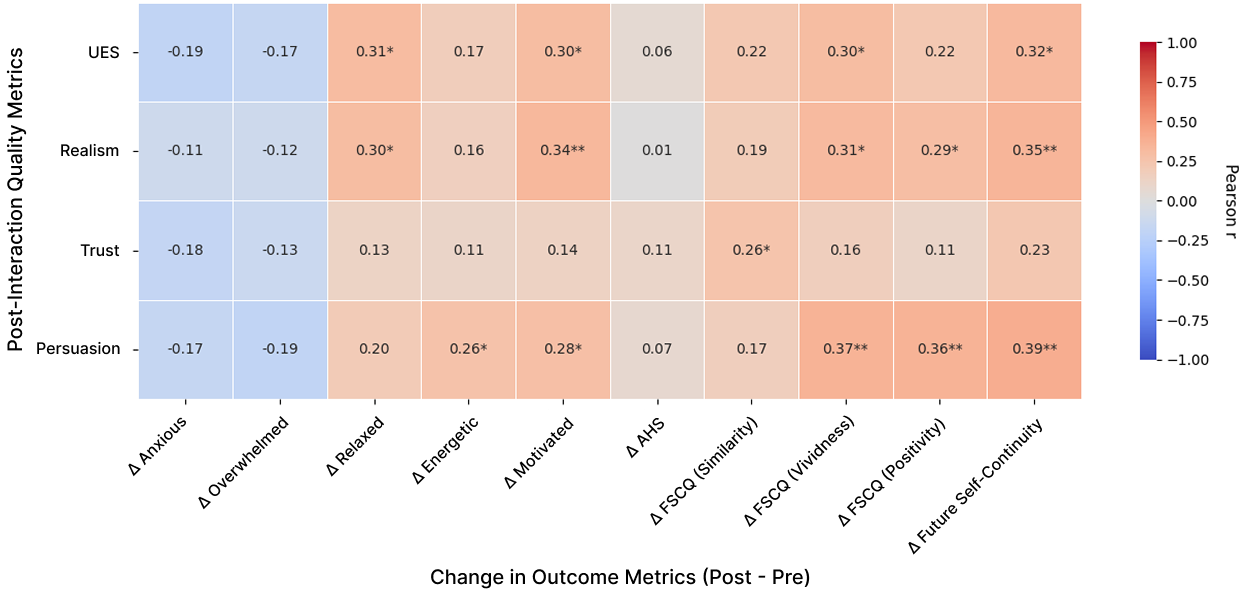}
    \caption{
        Pearson correlations between post-interaction quality metrics and change scores (POST–PRE), corrected for false discovery rate (FDR; *q~<~0.05, **q~<~0.01, ***q~<~0.001). 
        Positive correlations (red) indicate that higher interaction quality was associated with greater improvements in outcomes. 
        All four interaction quality dimensions showed significant associations with improvements in future-self–related outcomes.
    }
    \label{fig:correlation}
\end{figure*}

Beyond between-condition differences, we examined whether interaction quality predicted affective gains. Pearson correlations with FDR correction revealed consistent associations between post-interaction quality metrics and change scores (Fig. \ref{fig:correlation}). User Engagement Scale positively correlated with improvements in relaxation (r = 0.31, q < 0.05), energy (r = 0.30, q < 0.05), FSCQ Vividness (r = 0.30, q < 0.05), and Future Self-Continuity (r = 0.32, q < 0.05).

Perceived Realism showed positive correlations with changes in relaxation (r = 0.30, q < 0.05), energy (r = 0.34, q < 0.01), FSCQ Vividness (r = 0.31, q < 0.05), FSCQ Positivity (r = 0.29, q < 0.05), and Future Self-Continuity (r = 0.35, q < 0.01). Trust positively correlated with improvements in FSCQ Similarity (r = 0.26, q < 0.05). Persuasion demonstrated the strongest associations, correlating with changes in energy (r = 0.26, q < 0.05), motivation (r = 0.28, q < 0.05), FSCQ Vividness (r = 0.37, q < 0.01), FSCQ Positivity (r = 0.36, q < 0.01), and Future Self-Continuity (r = 0.39, q < 0.01). These findings suggest that perceived interaction quality, particularly persuasiveness and realism, serve as meaningful predictors of intervention efficacy.

\subsection{Modality Shapes Conversational Themes}

\begin{figure*}
    \centering
    \includegraphics[width=1\linewidth]{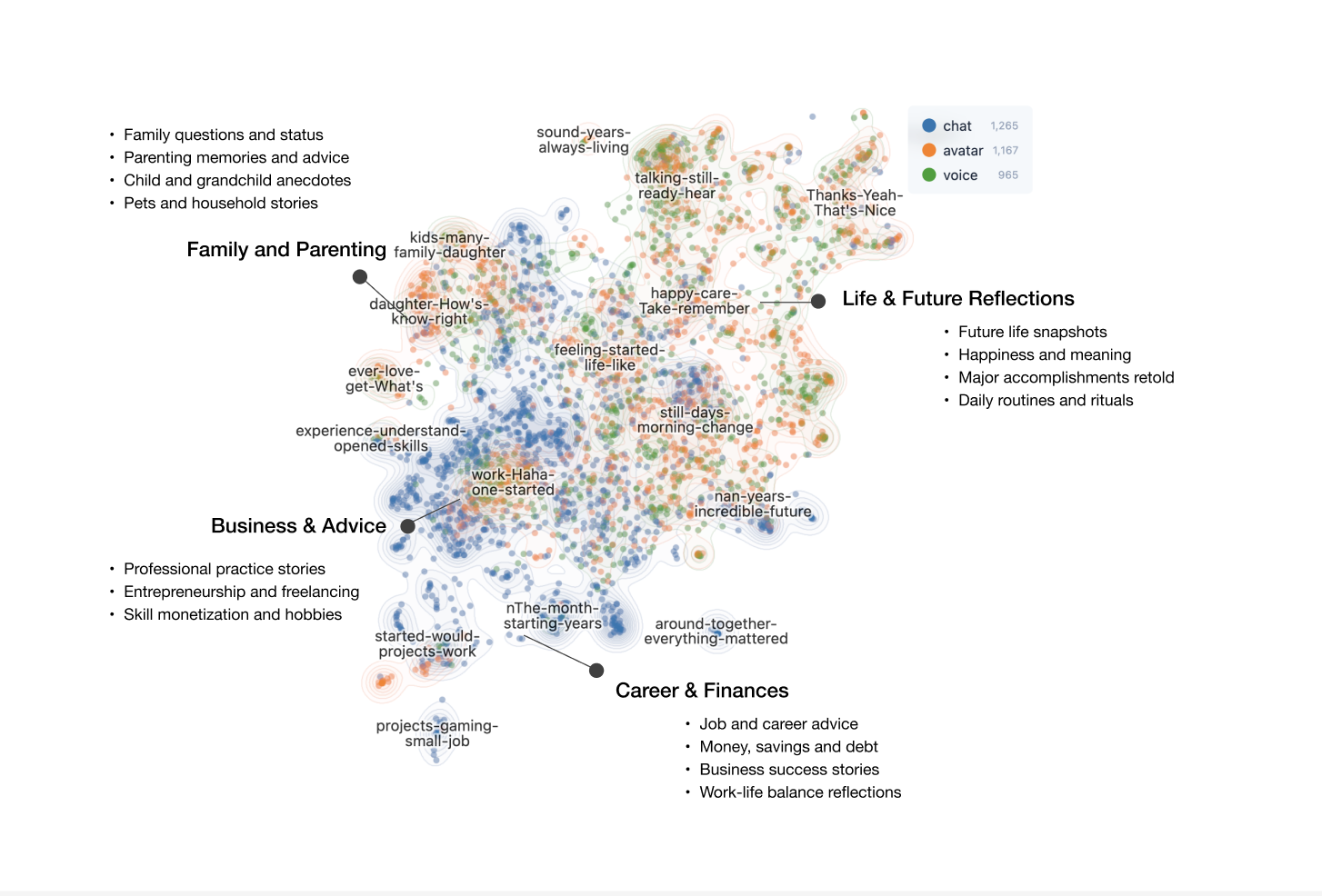}
    \caption{This figure visualizes the results of the correlation analysis, examining the relationships between the perceived realism of the future-self interaction and various outcomes.}
    \label{fig:conversational-patterns}
\end{figure*}

To identify thematic patterns in user-AI conversations, we performed hierarchical clustering analysis on conversation content (Fig. \ref{fig:conversational-patterns}). The analysis revealed four primary thematic clusters: \textit{Family \& Parenting} encompassed conversations about family relationships, parenting experiences and advice, anecdotes about children and grandchildren, and household life including pets. \textit{Career \& Finances} focused on professional development and career guidance, financial management including savings and debt, business achievements, and work-life balance considerations. \textit{Business \& Advice} emphasized professional practice narratives, entrepreneurship and freelancing experiences, monetization of skills and hobbies, and career growth strategies. \textit{Life \& Future Reflections} explored envisioned future scenarios, discussions of happiness and life meaning, recollections of major accomplishments, and descriptions of daily routines and rituals. 

Cluster distribution varied substantially across modalities. In the chat condition (\textit{n}(messages)=1,265), \textit{Career, Business \& Advice} dominated (52.6\%), followed by \textit{Family \& Parenting} (19.8\%) and \textit{Life \& Future Reflections} (18.7\%). The avatar (\textit{n}=1,167) and voice (\textit{n}=965) conditions showed more balanced distributions, with \textit{Life \& Future Reflections} most prevalent (31.1\% and 26.0\% respectively), followed by \textit{Family \& Parenting} (20.5\% and 18.9\%) (Fig. \ref{fig:conversational-patterns}). These patterns demonstrate that text-based interactions emphasized practical career concerns while embodied modalities facilitated deeper exploration of existential considerations including meaning, identity, and future aspirations.

\section{Discussion}

This study demonstrates that AI-generated future-self interactions reliably strengthen Future Self-Continuity—particularly vividness, positivity, and overall continuity—and improve hope and motivational states across all intervention modalities. Additionally, our systematic model evaluation established Claude 4 as the superior foundational model, outperforming competing LLMs by 40.8\% across psychological and interaction dimensions. Building on this optimized conversational AI foundation, all three interventions significantly increased hope (avatar: +0.37, p < .001; text: +0.33, p = .009; voice: +0.37, p = .012), with voice and text also enhancing motivation and energetic levels, while text reduced feelings of being overwhelmed. However, no between-condition differences emerged after controlling for baseline scores, indicating that psychological benefits stem from personalized future-self conversation rather than specific presentation formats. Although avatar produced the numerically largest increase in future-self vividness (B = 1.08), all three personalized modalities showed comparable improvements on most outcomes compared to control. Realism was the only interaction-quality measure revealing significant between-condition differences, with text and voice rated higher than control. Critically, interaction quality metrics—particularly persuasiveness (r = 0.39), realism (r = 0.35), and user engagement (r = 0.32)—strongly predicted gains in Future Self-Continuity and motivation across modalities. Conversational content analysis revealed that modality shaped reflection themes: text emphasized career planning (52.6\%), while voice and avatar facilitated existential and relational exploration (31.1\% and 26.0\% for life reflections). Together, these findings indicate that psychological impact derives from conversational personalization and perceived interaction quality rather than visual embodiment, demonstrating that scalable conversational systems can effectively support future-oriented thinking without requiring avatar interfaces, which can be more resource-intensive to implement.

\subsection{Personalization Drives Future Self-Continuity Gains Across All Modalities}

All personalized modalities—text, voice and avatar—significantly strengthened Future Self-Continuity (FSC) compared to control. After controlling for baseline scores, voice showed the largest effect (B = 0.76, SE = 0.20, q = 0.004), followed by avatar (B = 0.74, SE = 0.24, q = 0.016) and text (B = 0.60, SE = 0.21, q = 0.036), yet no intervention modalities differed significantly from each other. These findings align with research demonstrating that personalized future-self interventions \cite{pataranutaporn2024future,dechant2025future} and reflective engagement with one's future self \cite{Chishima2021TemporalDD,Chishima2020ConversationWA} enhance self-continuity. Our results extend this evidence by showing that AI-generated future selves yield comparable improvements across different modalities. The robust FSC gains observed across all conditions are consistent with established research linking future self-connection to improved decision-making \cite{ErsnerHershfield2009DontST}, reduced delinquency \cite{vanGelder2013VividnessOT}, enhanced academic performance \cite{adelman2017feeling}, better health behaviors \cite{rutchick2018future}, and greater meaning in life \cite{hong2024future}. Moreover, our findings extend recent work on AI-generated characters for personalized interventions \cite{pataranutaporn2021ai,jeon2025letters}, demonstrating that conversational AI can effectively facilitate future-self dialogue across multiple presentation formats while maintaining psychological and affective benefits.

Examining FSC subscales reveals nuanced patterns. For vividness, all three intervention conditions significantly outperformed control, with avatar showing the largest adjusted effect (B = 1.08, SE = 0.27, q = 0.003), followed by voice (B = 0.88, SE = 0.26, q = 0.008) and text (B = 0.86, SE = 0.25, q = 0.008). Within-condition pre-post comparisons reinforced this pattern: avatar (+1.36; t = 5.58, p < .001), text (+1.12; t = 5.11, p < .001), and voice (+0.96; t = 4.56, p < .001), while control showed no significant change (p = .070). The avatar condition's advantage in vividness aligns with prior work showing that visual representations enhance self-continuity through helping people perceive and imagine more concretely \cite{van2022interaction,Ganschow2021LookingBF,hershfield2011future}. This visual concreteness has been explored to reduce psychological distance more effectively than purely verbal or auditory cues \cite{vanGelder2013VividnessOT}, making the future self feel more tangible. Recent advances in AI-driven talking head synthesis \cite{sun2025vividtalk,kim2025moditalker} and virtual human technologies \cite{cui2023virtual,chen2024generative} have made such photorealistic representations of the future-self increasingly accessible. However, the modest numerical advantage of avatars over text (a difference of only 0.22 in adjusted effects) suggests that while visual embodiment contributes to vividness, conversational personalization and narrative coherence are equally critical \cite{neururer2018perceptions,huang2022perceived}.

Crucially, no significant differences emerged between modes for overall FSC or any subscale. While all three personalized conditions significantly outperformed control, they did not differ significantly from each other, indicating that  autobiographically grounded simulation of one's future self is sufficient to strengthen psychological connectedness to the future, even without high-fidelity perceptual embodiment. This finding diverges from research predicting that higher-fidelity modalities should enhance psychological outcomes through increased social presence and embodiment \cite{Ochs2022MultimodalBC,Haresamudram2024TalkingBT}. Instead, our results support the Self-AI Integration framework \cite{alabed2022ai}, in which alignment with personal values and self-concept transcends interface characteristics. \cite{Chishima2020ConversationWA,Chishima2021TemporalDD,pataranutaporn2024future}. 

Hence, a key contribution of this study is the demonstration that resource-intensive avatar systems are not required to meaningfully strengthen future self-continuity when personalization and temporal coherence are maintained through high-quality conversational AI. This finding broadens the feasible design space for future-self systems—from high-fidelity virtual characters to lightweight conversational agents deployable on mobile devices—enabling scalable and equitable applications in domains such as financial decision-making and youth career development.

\subsection{Affective Benefits Emerge Without Modality-Specific Advantages}

Beyond FSC improvements, all three intervention conditions produced significant within-condition gains in hope: avatar (+0.37; t = 4.94, p < .001), text (+0.33; t = 2.89, p = .009), and voice (+0.37; t = 2.72, p = .012), while control showed no change (p = .183). Similarly, motivation increased significantly in text (Mpre = 5.00; Mpost = 5.52; t = 2.23, p = .036) and voice (Mpre = 5.04; Mpost = 5.61; t = 2.51, p = .020) conditions. Voice also increased energetic levels (Mpre = 4.48; Mpost = 5.00; t = 2.31, p = .030), while text reduced feelings of being overwhelmed (Mpre = 2.61; Mpost = 2.00; t = -2.08, p = .050).

However, after controlling for baseline scores, no significant between-condition differences emerged for hope, emotional states, or motivation (all q > .05 after FDR correction). This modality-independence suggests that engaging in structured reflection on one's future goals—regardless of presentation format—can buffer negative affect and activate goal-oriented thinking.

The absence of modality-specific advantages contrasts with research demonstrating that higher-fidelity modalities enhance emotional engagement through prosodic cues and social presence \cite{Oh2018ASR,reicherts2022s}. Two factors may explain this pattern. First, as conversational AI becomes increasingly prevalent, users may be more accustomed to discussing personal topics with chatbots, potentially mitigating the novelty effects observed in earlier interventions \cite{pataranutaporn2024future}. Second, while all conditions achieved relatively high realism scores (M = 5.41-5.71 for interventions vs. M = 4.60 for control), this elevated realism may have triggered uncanny valley effects when paired with behavioral inconsistencies, attenuating expected emotional benefits from higher-fidelity modalities \cite{pizzi2019virtual,katsyri2015review,hepperle2022aspects}.

Despite only modest affective gains, the consistent psychological benefits observed across all modalities demonstrate that embodiment is not a prerequisite for meaningful impact \cite{Oh2018ASR,schuetzler2018influence}. This underscores future-self communication as a mechanism grounded in reflection and temporal perspective-taking rather than sensory immersion \cite{hershfield2011future, Chishima2021TemporalDD,Chishima2020ConversationWA}. 

\subsection{Interaction Quality Predicts Outcomes More Strongly Than Modality}

While FSC and affective outcomes showed no between-modality differences, interaction quality metrics varied minimally across formats and emerged as robust predictors of intervention effectiveness. User engagement, trust, and persuasion did not significantly differ between avatar, voice, and text conditions (all F(3, 88) < 1.58, all p > .10). Only perceived realism showed a significant main effect (F(3, 88) = 4.17, p = 0.008), with text (M = 5.71, SD = 0.83) and voice (M = 5.55, SD = 1.15) rated significantly higher than control (M = 4.60, SD = 1.28; text vs. control: p < 0.01; voice vs. control: p < 0.05). Notably, avatar (M = 5.41, SD = 1.32) did not differ significantly from control after correction for multiple comparisons. This pattern indicates that the separation in realism reflected the presence versus absence of personalized future-self conversation rather than the superiority of any specific modality.

This finding challenges assumptions in embodiment research, suggesting that interaction quality depends more on conversational coherence and behavioral authenticity than on visual fidelity or vocal synthesis \cite{neururer2018perceptions,huang2022perceived}. The integrated self-memory architecture powered by Claude 4, which provided personalized responses grounded in participants' autobiographical details across all modalities, appears to have been the primary driver of perceived authenticity. The 40.8\% improvement achieved by migrating from ChatGPT-3.5 to Claude 4 underscores that model quality—spanning emotional regulation (M = 4.80-6.60), future self-connection (FSCQ Composite: M = 6.73), and interaction quality (Trust: M = 6.00, Persuasion: M = 5.80, Realism: M = 5.40, Human Likeness: M = 7.00), serves as an additional determinant of intervention effectiveness rather than modality choice.

Perceived realism showed positive correlations with changes in relaxation (r = 0.30, q < 0.05), energy (r = 0.34, q < 0.01), FSCQ Vividness (r = 0.31, q < 0.05), FSCQ Positivity (r = 0.29, q < 0.05), and Future Self-Continuity (r = 0.35, q < 0.01). These associations suggest that perceived authenticity—achieved through personalized, autobiographically grounded conversation—serves as a foundation for strengthening psychological connection to one's future self across all presentation formats. This aligns with research demonstrating that perceived authenticity in conversational agents depends on behavioral consistency and conversational coherence rather than visual fidelity \cite{neururer2018perceptions,huang2022perceived}, and that trust in AI systems emerges from reliable, contextually appropriate responses rather than anthropomorphic appearance \cite{Glikson2020HumanTI,Jacovi2020FormalizingTI}. When AI systems respond in ways that feel genuine and personally relevant, users naturally engage with them as social actors \cite{reeves1996media}, triggering psychological processes that strengthen future self-connection \cite{kirk2025human}. Notably, realism predicted energetic arousal (r = 0.36, p < .001) but reduced relaxation (r = -0.31, p = .001), suggesting that authentic future-self interactions demand active cognitive engagement and emotional investment rather than passive reception \cite{adam2025generating}. Effective future-self interventions require participants to mentally simulate and emotionally connect with their imagined futures—a process facilitated by conversational realism that makes the future self feel like a credible interlocutor \cite{Pataranutaporn2023LivingMA,pataranutaporn2021ai}.

User engagement and persuasion emerged as the strongest predictors of intervention effectiveness. Engagement correlated significantly with relaxation (r = 0.31, q < 0.05), energy (r = 0.30, q < 0.05), FSCQ Vividness (r = 0.30, q < 0.05), and Future Self-Continuity (r = 0.32, q < 0.05), while persuasion demonstrated the strongest associations: energy (r = 0.26, q < 0.05), motivation (r = 0.28, q < 0.05), FSCQ Vividness (r = 0.37, q < 0.01), FSCQ Positivity (r = 0.36, q < 0.01), and Future Self-Continuity (r = 0.39, q < 0.01). This pattern reveals a critical insight: the affective value of future-self interventions depends not merely on experiencing the interaction, but on how deeply users psychologically invest in it. Effective interventions require users to find the interaction compelling enough to genuinely consider the future self's perspective \cite{Matz2024ThePO,matz2024potential}. 

The strong relationship between persuasion and outcomes raises important questions about agency and autonomy in AI-mediated self-reflection \cite{Hancock2020AIMediatedCD,laitinen2021ai}: when does persuasive conversation facilitate authentic self-exploration versus algorithmic manipulation of identity narratives \cite{Jakesch2023CoWritingWO,Anderson2024HomogenizationEO}. Trust positively correlated with improvements in FSC Q Similarity (r = 0.26, q < 0.05), suggesting that users must believe the future-self representation is credible before internalizing its perspective \cite{de2016almost,Glikson2020HumanTI}. 

Importantly, these quality metrics operated within conditions— participants who felt more engaged and persuaded benefited more— yet these perceptions did not significantly differ between modalities (except realism). This indicates that how compelling the interaction feels matters far more than what form it takes, and that high-quality conversational AI can achieve persuasive engagement regardless of embodiment level.

\subsection{Modality Shapes Conversational Content and Cognitive Affordances}

While modalities showed equivalent psychological and affective outcomes, conversational content analysis revealed systematic thematic differences, indicating distinct cognitive affordances across presentation formats. Hierarchical clustering analysis demonstrated that text-based interactions emphasized instrumental career concerns (Career, Business \& Advice: 52.6\%), whereas avatar (31.1\%) and voice (26.0\%) conditions displayed more distributed thematic profiles with elevated representation of existential considerations (Life \& Future Reflections) and relational topics (Family \& Parenting: 20.5\% and 18.9\% respectively). 

These distributional differences support a cognitive fit hypothesis: text-based interfaces facilitate structured narrative construction and goal articulation, enabling the externalization of abstract aspirations into concrete objectives through mental contrasting \cite{abootorabi2025generative}. In contrast, embodied modalities lower barriers to affective exploration and personal discussions. Voice interactions, with their prosodic cues and conversational immediacy \cite{reicherts2022s}, evoke greater emotional intimacy \cite{Oh2018ASR}, while avatar-based interactions enhance social realism through visual presence \cite{nowak2001understanding,latoschik2017effect,waltemate2018impact}. This aligns with research showing that anthropomorphic modalities elicit stronger social-emotional responses \cite{reeves1996media}, activating relational schemas and shifting focus from instrumental planning toward identity-oriented themes \cite{oker2022embodied,schuetzler2018influence}, whereas text-based communication promotes analytical processing \cite{Fang2025HowAA,bailenson2001equilibrium}.

The observed thematic divergence suggests that modality selection should align with intervention objectives. Text-based formats are optimal for structured goal-setting, career planning, and financial decision-making where participants need to articulate specific action steps \cite{Sims2020TheFI,ErsnerHershfield2009DontST}. Conversely, voice and avatar modalities better support emotional processing, identity exploration, and other contexts where affective engagement and existential reflection are primary goals \cite{dechant2025future,pataranutaporn2024future}. For comprehensive interventions targeting both domains, hybrid approaches combining modalities may optimize outcomes by leveraging the distinct affordances of each format.

\subsection{Ethical Considerations and Design Implications}

These findings raise important ethical considerations for future-self system design. While realism and engagement enable psychological closeness and motivation—outcomes achievable across all modalities through high-quality conversational AI and personalization—the pursuit of hyperrealism carries potential risks. As AI systems achieve greater perceptual and behavioral fidelity, they may increasingly co-construct users' self-narratives and emotional responses, potentially blurring boundaries of personal agency. Hyperrealistic self-avatars may trigger uncanny valley discomfort \cite{Mori2012TheUV,weisman2021face} or risk shifting narrative authorship from user to algorithm. Maintaining clear boundaries between reflection and persuasion—preserving users' sense that their imagined future emerges from authentic self-reflection rather than algorithmic influence—remains essential for ethical deployment \cite{Hancock2020AIMediatedCD}.

These findings reposition embodiment not as a default path to improvement but as one design dimension among several, valued primarily when aligned with the type of self-reflection a context demands. Systems aiming to influence behavior should prioritize model quality, personalization, realism, and engagement—levers that can be achieved at any fidelity level with sufficiently advanced conversational AI—to create impactful and scalable future-self interventions.

\section{Limitations and Future Directions}

The present study has several limitations that inform future research directions. First, our single-session design examining immediate pre-to-post changes cannot assess whether observed effects persist over time or how repeated interactions shape long-term behavioral change. Longitudinal field studies are needed to examine how trust, identification, and motivation evolve over extended engagement with future-oriented AI agents.

Second, our sample of 92 participants (23 per condition) provides adequate power for detecting medium-to-large effects but may be underpowered for identifying subtle modality interactions or moderating effects of individual differences. The limited sample size also restricts generalizability across demographic groups, as our U.S.-based sample may not reflect how future-self interactions function across diverse cultural contexts where self-concept, temporal orientation, and attitudes toward AI vary systematically.

Third, technical constraints limited the realism and immersion of our implementations. Participants reported lag and desynchronization issues, as the lip-syncing model relied on pre-generated mouth movements and conversation quality depended on internet connection stability. These technical limitations are consistent with prior findings that latency reduces perceived authenticity and trust \cite{katsyri2015review}. Future implementations should leverage advances in real-time lip-sync models and higher-fidelity age-progression algorithms to minimize technical artifacts.

Fourth, the generic chatbot control may not have provided an ideal neutral baseline. Future research should employ modality-matched control conditions with structured conversational scaffolding to better isolate the effects of personalization versus presentation format.

\section{Conclusion}

"This experience brought me a totally new experience that I wasn't expecting. It was really cool to be able to relate to an older me and to talk to an older me. It brought me a glimpse of what the future may be like for me." This participant's reflection captures the profound impact of AI-generated future-self interactions, demonstrating that accessible conversational systems can meaningfully strengthen connections to one's future self.

Our findings reveal three key insights: (1) Claude 4 substantially outperformed competing LLMs (40.8\% improvement), establishing conversational AI quality as more critical than modality choice, while all personalized formats—text, voice, and avatar—significantly enhanced Future Self-Continuity, hope, and motivation relative to control with no significant differences between interventions; (2) Interaction quality metrics, particularly persuasiveness (r = 0.39 with FSC), realism, and engagement, emerged as robust predictors of affective gains operating within conditions—participants who found interactions more compelling benefited more regardless of format; (3) Modality systematically shaped conversational content, with text emphasizing career planning (52.6\%) and voice/avatar facilitating existential reflection (31.1\%, 26.0\%), revealing distinct cognitive affordances suited to different objectives.

These findings demonstrate that effective future-self interventions need not rely on resource-intensive technologies—high-quality conversational AI achieves meaningful psychological impact across formats when optimizing model quality, personalization, and behavioral authenticity. This modality-independence enables scalable deployment: text for structured goal-setting, voice and avatar for emotional processing, or hybrid approaches for comprehensive interventions.

The implications extend beyond future-self agents, offering guidance for AI-mediated self-reflection more broadly. As conversational AI achieves greater persuasive capacity, design decisions must balance psychological and affective benefits with ethical considerations around agency and narrative authorship. The strong association between persuasion and outcomes highlights both the power and responsibility in AI systems shaping users' self-narratives, underscoring the need to preserve authentic introspection over algorithmic influence.

\bibliographystyle{ACM-Reference-Format}
\bibliography{references}

\end{document}